# Study Of Spatial Biological Systems Using a Graphical User Interface

*Nigel J. Burroughs, George D. Tsibidis*

Mathematics Institute,
University of Warwick, Coventry,
CV47AL, UK
{njb,tsibidis}@maths.warwick.ac.uk

*William Gaze, Liz Wellington*

Department of Biology,
University of Warwick, Coventry,
CV47AL, UK
{w.gaze,e.m.h.wellington@warwick.ac.uk}

**Abstract**

In this paper, we describe a Graphical User Interface (GUI) designed to manage large quantities of image data of a biological system. After setting the design requirements for the system, we developed an ecology quantification GUI that assists biologists in analysing data. We focus on the main features of the interface and we present the results and an evaluation of the system. Finally, we provide some directions for some future work.

## 1. Introduction

Biologists experience difficulties in managing and analysing large amounts of data. To tackle this problem, we designed a GUI, which incorporates tools for a systematic analysis of image data. The objective of the interface is to enable biologists to manage and analyse a large number of images by aiding classification of, and viewing of object groups that are present in image set. Its design is based on a consideration of system requirements and an assessment was conducted by allowing experts to evaluate the system. This paper focuses on three main aspects of the development of the GUI:
- The requirements of the system design.
- The design and development of the interface.
- An evaluation of the system focusing on the quality of the results.

## 2. System requirements

The design of the system should be based on the following requirements: The interface should
- offer a rapid data overview
- require minimum intervention by the user for object classification,
- allow fast error identification,
- be dynamic and adaptive to changes,
- be reliable and robust.

## 3. Application and User Interface Design

A biology experiment was carried out in the Department of Biology at the University of Warwick whose objective was the study of the ecology of a biological system (*Acanthamoeba polyphaga*, a ubiquitous unicellar amoeba, and *Salmonella typhimurium*, an enteric bacterial pathogen on non nutrient on agar). A large number of images were obtained (a series of at least 3,000 spatial images, each image about 2MB, 1022 x1024) and within each image a number of biological objects were present (see Figure 1). The aim of the analysis is to extract and classify all objects in the data set, distinguish dead from live objects, group objects into classification categories and finally carry out a detailed quantitative and qualitative spatial analysis of the results.

Basic Image analysis techniques (Weeks, 1996) were used and modified to locate and count all objects. The refractive halo was used to locate all objects in each image by intensity thresholding, giving morphological information about the objects such as perimeter, centre of mass and shape details. To distinguish dead from live objects, a delayed image (4secs) comparison (image subtraction) was performed.

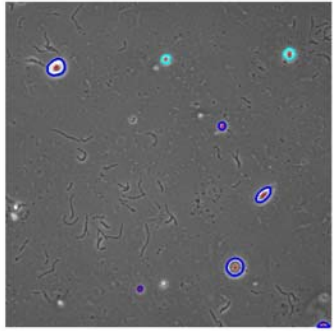

**Figure 1:** Typical image containing amoeba and cysts
(perimeter and centre of mass shown) and bacteria, e.g. filaments.

However, four problems are encountered with the automated procedure:
- Classification error,
- Incorrect perimeter,
- Missing objects,
- Multiple objects (refractive halo objects comprising more that one biological object)

To overcome these problems, we built a GUI (see Figure 2) using Matlab 6, which is one of a pair of graphical interfaces we have developed, an amoeba classification GUI (this paper) and a bacterial quantification GUI (not shown). The amoeba classification GUI allows the user to intervene and correct interactively classification and morphological errors.

There exist two types of classification for the objects of the experiment, one is based on the type of the object (amoeba, cyst, etc) while the second is based on the state of the object (live or dead). The GUI can help the user to avoid a time consuming manual classification for every individual object by running an automated classification procedure. The basis of these algorithms is illustrated in the two plots in Figure 3; cysts and dead amoebae are clustered in the lower section of the respective graphs and can thereby be classified based on approximate clustering. As a result, the interface offers the facility to create clustering patterns that allow a faster classification and decision making process. This is very important in minimising the effort of the user to achieve object classification throughout the entire data set.

The GUI consists of viewing facilities and modification and analysis tools. It operates as a management tool allowing the user to view summary information for individual objects or whole categories of objects, and also conduct a detailed analysis. The modification tools enable the user to either redraw the perimeter of an object if it is incorrect, introduce new objects if they were missed by the automated image analysis and to modify an object's classification at any time. All these manual changes are recorded and an overall analysis is adapted to the changes making the GUI a dynamic interface.

The analysis tools are used to facilitate a fast quantitative and qualitative analysis for objects belonging to a particular category by enabling the user to view graphical representations of the results in terms of plots (see Figure 3).

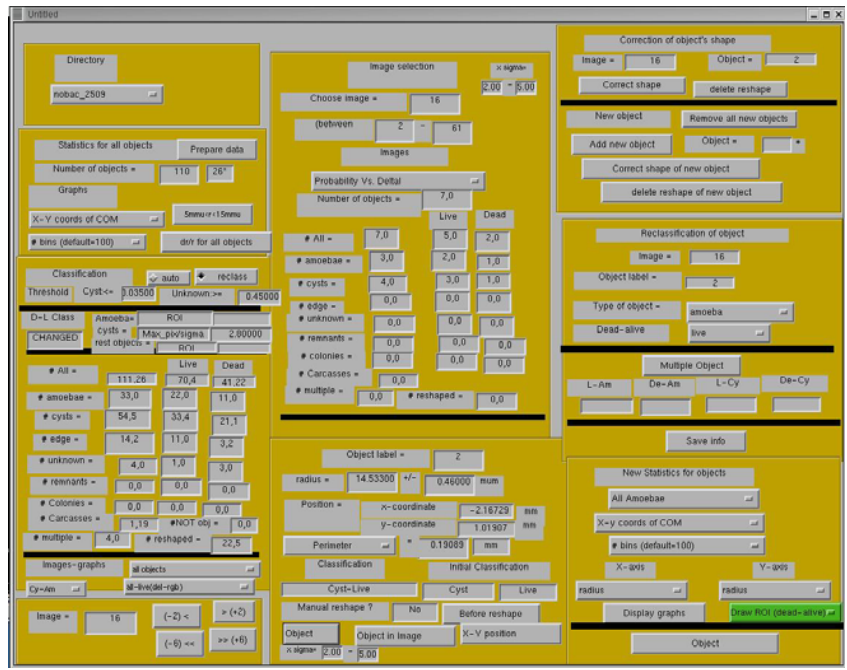

**Figure 2:** The Graphical User Interface

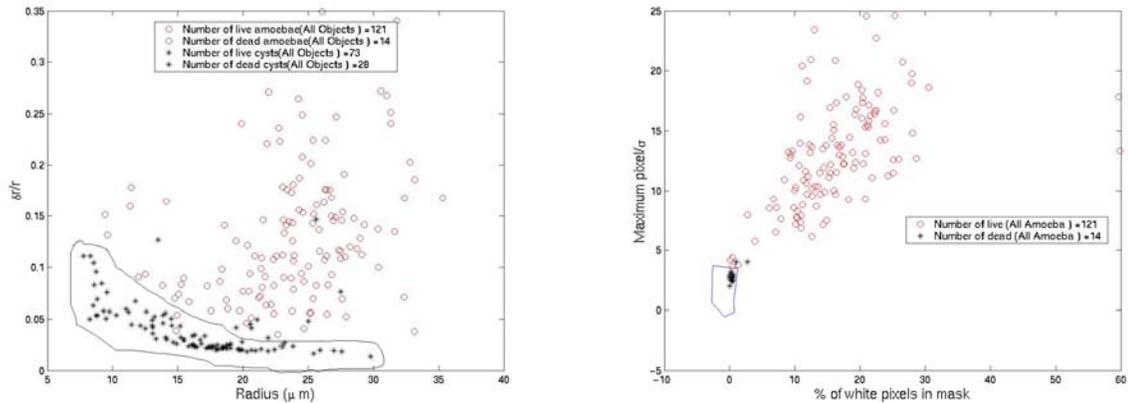

**Figure 3:** Amoeba-cyst and Dead-live classification. *Left panel*: shape analysis
using relative standard deviation of object by radius. *Right panel*: maximum pixel intensity
in object (relative to an estimated intensity standard deviation $\sigma$) by fraction of live
pixels in object (live being significantly above noise, 95%).

The viewing facilities aids the user to scan quickly a library of objects belonging to each category (see Figure 4), compare them and decide whether a object classification change is required. The decision of the user can be assisted by the fact he/she is able to compare all objects that are in the same category and determine an optimal object classification criterion. If the classification of an object has been modified, it is indicated to inform the user of the change. In the dead or live object category, a library of images resulting from the delayed image comparison for all objects is produced and it provides a clear and fast view of the dead-live classification.

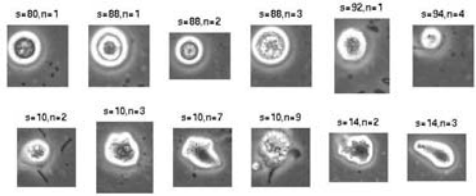

**Figure 4:** Library of images (upper row: cysts, lower row: amoeba)

The user can also employ the viewing tools to obtain data, measurements, confidence statistics and images for individual objects. In Figure 5, the first row shows the image of an extracted object (left box) and the same figure with the perimeter of the object drawn (right box) while the second row provides a summary of measurement and results (left box) and the delayed image subtraction for live or dead analysis (right box).

Another useful feature of the viewing facility is that it allows to view the full image sequence displaying all existing objects and their perimeters and centres of mass. The advantage of supplying a tool that allows to go through the whole set of images and see this type of characteristics provides a fast process of determining which objects will be used in the analysis and which objects must be included and redrawn manually.

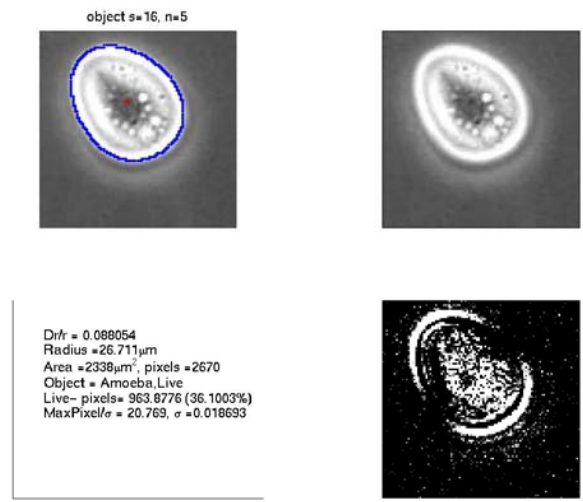

**Figure 5**: Images and measurements for an individual object

Both analysis and viewing facilities are very helpful tools because they simplify data management and reduce the number of objects requiring manual reclassification. This is a very important issue in terms of speeding up the analysis of a huge data set.

Finally, the efficiency of the interface can be tested by comparing the results derived from the automated and manual classifications of the objects and measuring the level of accuracy and confidence level of the automated procedure.

## 4. Evaluation

The interface was run on a Sun (Unix) platform with 1Gb of memory. It is mobile in the sense that it can run on Windows or Mac platforms provided there is substantial memory and the computer is equipped with a drive that can accommodate a large number of data.

The main objective of the evaluation procedure was to assess the set of requirements against the decision of experts. Two biologists from the Department of Biology at the University of Warwick helped the evaluation of the system. The focus was centred on the efficiency of the GUI and the simplicity it provided in the analysis of the data:

- The accuracy in the object classification was very high; More than 85% of the objects in the dataset did not require manual classification.
- The number of the objects that required their shape to be redrawn or had to be introduced manually was very small ($\approx$5% of the total number of objects).
- The pattern recognition for the objects through the ROI clustering led to a faster object classification.
- The GUI is user friendly.

These demonstrate that the GUI leads to an efficient fast classification, minimises the need for the user intervention and simplifies the analysis.

## 5. Conclusions-Future Work

This GUI helped to analyse data collected in the Department of Biology at the University of Warwick. The aim of this project was primarily the simplification of object analysis by providing the means to assist a researcher in analysing a biology experiment in a fast and efficient way. The fundamental findings of our inquiry are that the GUI:

- operates as an information management tool,
- helps to group data and produce graphical representations that simplify the analysis,
- enables pattern recognition for object identification,
- minimises user intervention in object classification,
- is dynamic and adaptive to changes,
- is mobile provided the hardware requirements are satisfied

The present interface is one of a pair of GUIs (the other focuses on bacteria extraction and bacteria coverage in a huge set of images) that have been implemented with excellent results. An improvement would be to build an on-line version of the interface. A further study is ongoing that analyses sequences of images (videos) by studying trajectories of biological objects.